\documentclass[pra,reprint,superscriptaddress]{revtex4-1}

\usepackage{amsmath,amsfonts,amssymb,amsthm,braket,graphicx,color,lipsum}
\usepackage[colorlinks=true,citecolor=blue,linkcolor=magenta]{hyperref}
\usepackage{bm}

\usepackage{aligned-overset}

\usepackage[normalem]{ulem}
\usepackage[dvipsnames]{xcolor}

\global\long\def\ket#1{|#1\rangle}
\global\long\def\bra#1{\langle#1|}
\global\long\def\proj#1#2{|#1\rangle\langle#2|}
\global\long\def\inner#1#2{\langle#1|#2\rangle}
\global\long\def\tr{\mathrm{tr}}

\global\long\def\im{\imath}

\newcommand{\bG} {\bm{G}}
\newcommand{\bS} {\bm{\Sigma}}
\newcommand{\bGa} {\bm{\Gamma}}

\newcommand{\bHS}{\bm{\bar{H}}_\qs}

\renewcommand{\Im} {\operatorname{Im}}

\newcommand{\ql}{\mathcal{L}}
\newcommand{\qs}{\mathcal{S}}
\newcommand{\qr}{\mathcal{R}}

\newcommand{\ok}{\proj{v_k}{v_k}}
\newcommand{\bk}{\bra{v_k}}
\newcommand{\kk}{\ket{v_k}}
\newcommand{\ik}{\left| v_k \right|^2} 

\newcommand\trick[1]{}

\renewcommand{\[}{\begin{equation}}
\renewcommand{\]}{\end{equation}}

\begin{document}

\title{Comment on ``Quantum transport with electronic relaxation in electrodes: Landauer-type formulas derived from the driven Liouville--von Neumann approach'' [The Journal of Chemical Physics {\bf 153}, 044103 (2020)]}
\author{Michael Zwolak}
\email{mpz@nist.gov}
\affiliation{Biophysical and Biomedical Measurement Group, Microsystems and Nanotechnology Division, Physical Measurement Laboratory, National Institute of Standards and Technology, Gaithersburg, MD 20899, USA}

\maketitle

In a recent article, Chiang and Hsu [The Journal of Chemical Physics {\bf 153}, 044103 (2020)] examine one and two site electronic junctions {\em identically} connected to finite reservoirs~\cite{chiang_quantum_2020}. For these two examples, they derive analytical solutions, as well as provide asymptotic analyses, for the steady-state current from the driven Liouville--von Neumann (DLvN) equation~\cite{zelovich_state_2014} -- an open system approach to transport in non-interacting systems where relaxation maintains a bias. The two site junction they examine has destructive interference, which they show leads to slow convergence of the DLvN to the Landauer limit with respect to reservoir size and relaxation. 

We previously derived the general solution for the steady-state current in both the DLvN and its many-body analog [Gruss et al., Scientific Reports {\bf 6}, 24514 (2016)]~\cite{gruss_landauers_2016}. The many-body analog is a Lindblad master equation, which, when restricted to non-interacting systems, is exactly the DLvN~\cite{elenewski_communication_2017}. Here, we demonstrate that applying the more general expression from Ref.~\onlinecite{gruss_landauers_2016} to identical left and right reservoirs (i.e., finite reservoirs with the same density of states and coupling to the system) and Markovian relaxation provides a simple analytic form that applies to arbitrary, but identically connected, junctions. Moreover, Chiang and Hsu indicate that there is no proof of the convergence of the DLvN current to the Landauer limit, as well as indicate that prior asymptotic analyses are of a numerical nature. However, the current does limit to the Landauer and Meir-Wingreen result for non-interacting and interacting systems, respectively~\cite{gruss_landauers_2016,zwolak_analytic_2020}. Convergence occurs as the reservoirs' lesser Green's functions begin conforming to the fluctuation-dissipation theorem. Our approach sheds light on the behavior Chiang and Hsu observe for destructive interference. Finally, we show that the analytical results yield the asymptotic formulas derived in Ref.~\onlinecite{gruss_landauers_2016}.

When our prior, exact solution is applied to non-interacting junctions, it yields Eq.~(B13) of Ref.~\onlinecite{gruss_landauers_2016},
\[ \label{eq:nonintCurr_Sum}
I=e \int \frac{d\omega}{2\pi} \tr \left[ \tilde{\bGa}^\ql \bG^a \bGa^\qr \bG^r - \bGa^\ql \bG^r \tilde{\bGa}^\qr \bG^a \right] ,
\]
where $I$ is the steady-state current, $e$ is the electron charge, $\tilde{\bGa}^\alpha$ ($\bGa^\alpha$) are the weighted (unweighted) spectral density of the $\alpha=\ql$ and $\qr$ reservoirs, and $\bG^{r (a)}$ are the retarded (advanced) Green's functions of the system $\qs$. This equation is for arbitrary non-interacting systems, including reservoirs that are not identical in either density of states or coupling to the system. 

Chiang and Hsu assume, however, that the reservoirs are identically coupled to the system and have the same density of states. These are very strict conditions, and the former is essentially never satisfied in real junctions. For instance, even a two site junction with each site connected to one reservoir only, e.g., see the model in Ref.~\onlinecite{wojtowicz_open-system_2020}, does not satisfy the coupling condition (Chiang and Hsu's two site junction has one site connected to both reservoirs). Similarly, the paradigmatic benzene dithiol junction also does not, since the ends are bound to separate electrodes -- identical coupling would require that both thiol groups are bound to both electrodes. Nevertheless, it is important to explore all cases to learn more about the DLvN approach, as well as related techniques. 

Within Eq.~\eqref{eq:nonintCurr_Sum}, these two conditions are tantamount to $\bGa^\ql = \bGa^\qr$ (for finite systems this entails modes located at the same frequencies) and also yield that the weighted spectral densities are equal except for the Fermi-Dirac function weighing them. Under this assumption, the standard identity for non-interacting systems~\cite{haug_quantum_2008}
\[ \label{eq:nonintID}
\bG^r-\bG^a = -\im \bG^r (\bGa^\ql + \bGa^\qr) \bG^a = -\im \bG^a (\bGa^\ql + \bGa^\qr) \bG^r 
\]
yields the equations  
\[ \label{eq:nonintID_PC}
\bG^r-\bG^a = - 2 \im \bG^a \bGa^\qr \bG^r = - 2 \im \bG^a \bGa^\ql \bG^r .
\]
Employing this identity and the cyclic property of the trace in Eq.~\eqref{eq:nonintCurr_Sum} yields
\[ \label{eq:nonintCurr_PC}
I=\frac{\im e}{2} \int \frac{d\omega}{2\pi} \tr \left[ \left( \tilde{\bGa}^\ql - \tilde{\bGa}^\qr \right) \left( \bG^r-\bG^a \right) \right] .
\]
To go further, we need the form of the (difference in the) weighted spectral density
\[
\tilde{\bGa}^\ql - \tilde{\bGa}^\qr = \im \sum_{k\in \ql} \left(\tilde{f}_k^\ql - \tilde{f}_k^\qr \right) \left[g_{k}^{r}(\omega) - g_{k}^{a}(\omega) \right] \ok ,
\]
where $\inner{i}{v_k}=v_{ik}$ is the coupling between $i\in\qs$ and $k\in\ql$ (or $\qr$), the $g_{k}^{r(a)}=1/(\omega-\omega_k \pm \im \gamma_k/2)$ are the ``isolated'' retarded (advanced) Green's functions for $k\in\ql$ (or $\qr$) with relaxation $\gamma_k>0$, and $\tilde{f}_k^{\ql(\qr)}$ are the Fermi-Dirac occupations evaluated at frequency $\omega_k$ and bias $\mu_{\ql(\qr)}$. The sum is over only $k \in \ql$ since $\ql$ and $\qr$ are identical.

One then performs integrals in the upper or lower complex plane depending on analyticity of the integrand. Since we do not know {\em a priori} the poles of $\bG^{r(a)}$, we make sure to integrate in the upper (lower) plane when $\bG^r$ ($\bG^a$) appears since they are analytic there. This yields
\[ \label{eq:Curr_NI_PC}
I = - e \sum_{k\in \ql} \left(\tilde{f}_k^\ql - \tilde{f}_k^\qr \right) \bk \Im {\bG^r (\omega_k + \im \gamma_k/2 )} \kk ,
\]
where $\bG^{r}(\omega) = 1/(\omega - \bHS - \bS^{r})$, the self-energy is $\bS^{r}=\sum_{k\in\ql\qr} g^{r}_k \ok = 2 \bS^{r}_\ql$, and $\bHS$ is the single-particle Hamiltonian of $\qs$~\footnote{The continuum limit of Eq.~\eqref{eq:Curr_NI_PC} can be easily taken, resulting in an integral over the reservoir bandwidth.}. We emphasize that these are $N_\qs \times N_\qs$ operators where $N_\qs$ is the number of junction modes, whereas Chiang and Hsu have $N_r \times N_r$ operators where $N_r$ is the number of reservoir modes.

Equation~\eqref{eq:Curr_NI_PC} yields Chiang and Hsu's results for their two examples, their Eqs.~(13) and (30), after some simple manipulation. However, it applies to arbitrary non-interacting junctions/systems $\qs$ instead of two specific examples (but still requires identical reservoirs, both in system-reservoir coupling and in density of states) and allows for arbitrary, inhomogeneous $\gamma_k$'s. We generalize further in a companion article~\cite{zwolak_analytic_2020}. 

Moreover, Chiang and Hsu's claim that, ``there is no mathematical proof that the steady-state current of the DLvN approach is equivalent to the Landauer current," is incorrect. Already, in Ref.~\onlinecite{gruss_landauers_2016}, we showed the equivalence between the DLvN and the Landauer expression as $\gamma_k \to 0$ for {\em all} non-interacting systems~\footnote{Similarly, the Meir-Wingreen formula holds as $\gamma_k \to 0$.}. It relies only on noting that the quantities $\tilde{f}_k^{\ql(\qr)}=f_{\ql(\qr)}(\omega_k)$ are approximately equal to $f_{\ql(\qr)}(\omega)$ when weighted by the Lorentzian spectral density of a single reservoir mode at $\omega_k$. That is, in the weighted spectral density, for a single mode with relaxed but otherwise isolated $g_k^{r(a)}$, we have
\begin{align}
\lim_{\gamma_k \to 0} \im  \left[ g_{k}^{r}(\omega) - g_{k}^{a}(\omega) \right] \tilde{f}_k^\alpha & = 2 \pi \delta(\omega-\omega_k) \tilde{f}_k^\alpha \notag \\
& = 2 \pi \delta(\omega-\omega_k) f_\alpha(\omega) .
\end{align}
This correspondence permits replacing $\tilde{f}_k^\alpha=f_{\ql(\qr)}(\omega_k)$ with $f_{\ql(\qr)}(\omega)$ in the weighted spectral density, provided that the variation of the Fermi-Dirac distribution is small over the Lorentzian's width. This results in the Landauer formula (see Eq.~(B16) in Ref.~\onlinecite{gruss_landauers_2016}) -- not just a ``Landauer-type'' formula -- for {\em finite reservoirs}, which is approximate when the $\gamma_k$'s are small, but still non-zero. 
The zero relaxation limit can be taken provided that the reservoirs are infinite (if the reservoirs are not infinite, one obtains zero current in this limit), where Eq.~\eqref{eq:nonintCurr_Sum} goes to 
\[ \label{eq:nonintCurrStandard}
I = e \int \frac{d\omega}{2\pi} \left( f_\ql (\omega) - f_\qr (\omega) \right) \tr \left[ \bGa^\ql \bG^r \bGa^\qr \bG^a \right] ,
\]
with all quantities evaluated with $\gamma_k \to 0$. This limit is the normal limit in non-equilibrium Green's functions. Viewed in this light, the Lindblad master equation (which is identical to the DLvN for non-interacting systems) is a way to ensure causality and take the continuum limit (first the reservoir size to infinity and then $\gamma_k \to 0$; alternatively, $\gamma_k \to 0$ while holding the mode spacing $\ll \gamma_k$), for which stationary states still result for finite systems.

Note that, while this Landauer expression is in the limit of zero $\gamma_k$ for the Markovian relaxation, one can derive an {\em exact} Landauer expression for non-zero $\gamma_k$ when the relaxation is non-Markovian~\cite{gruss_landauers_2016}, and there will be a broad array of results of this type. That is, there is a well-defined transmission coefficient at finite relaxation provided that the relaxation is physical (Markovian relaxation is not, see below) and it arises from non-interacting source in the single-particle sense (e.g., an alternative situation where this arises, besides non-Markovian relaxation, is the Landauer-B\"uttiker approach to dephasing). 

This is in contrast to the Landauer-type formulas, Eq.~(25) for the single site junction and Eq.~(31) for the two-site junction in Chiang and Hsu. Those formulas look like the Landauer formula, but what they call the transmission is not a proper transmission function (it limits to one as $\gamma \to 0$). This is evident by the fact that there are full and empty states in the reservoirs beyond the band edges ($E_\mathrm{min (max)}$ in their terminology), e.g., see the spectral density above. Thus, since the transmission function gives the probability of electrons at some energy to transverse the junction, what Chiang and Hsu call the transmission can not be transmission as it is undefined for the model in regions where there is a non-zero density of states in the reservoirs (i.e., full and empty states smeared to outside of the bandwidth).

The Landauer correspondence is in many ways unsatisfying, as Chiang and Hsu's study of the two site model with destructive inference and other results~\cite{gruss_communication_2017} attest. It does not tell us how fast the result converges to the relaxation-free Landauer result. Additional results in Ref.~\onlinecite{gruss_landauers_2016} can help in that regard, although they do not provide the complete picture. There are Markovian and non-Markovian versions of relaxation, as noted above, with the latter having a proper Fermi level (i.e., the fluctuation-dissipation theorem holds) and where a Landauer formula is {\em always} obeyed for non-interacting junctions (see Eq.~(3) in Ref.~\onlinecite{gruss_landauers_2016}), as does the Meir-Wingreen formula (see Eq.~(A25) in Ref.~\onlinecite{gruss_landauers_2016} or Ref.~\onlinecite{zwolak_analytic_2020}). The Markovian and non-Markovian Green's functions, specifically the lesser Green's function $g^<_k$, are identical except the latter already contains $f_{\ql(\qr)}(\omega)$ instead of $f_{\ql(\qr)}(\omega_k)$ (the retarded and advanced Green's functions are identical). Thus, before converging to the Landauer expression with zero relaxation, one has to at least have convergence to the non-Markovian non-zero relaxation Landauer formula, which is controlled solely by convergence of $g^<_k$.

The integrated error in $g^<_k$, comparing Markovian and non-Markovian versions, with the one-norm of the difference is bounded by, see Eq.~(C3) in Ref.~\onlinecite{gruss_landauers_2016},
\[
\frac{\hbar \gamma_k}{4 k_B T} \ln \frac{k_B T}{\hbar \gamma_k} ,
\]
where $\hbar$ is the reduced Planck's constant, $k_B$ is Boltzmann's constant, and $T$ is the temperature. In words, convergence is slow (slower than linear) regardless of temperature and $\gamma_k$ should be quite small. The dominant contribution -- and the place where mode density and relaxation are the most important -- is at the Fermi level, as it is there that the Fermi-Dirac distribution changes the most rapidly (with slope $-1/4 k_B T$). 

Chiang and Hsu assume zero temperature, and thus one does not expect convergence to necessarily occur until $\gamma$ is zero, as there is no small parameter due to the infinite slope at the Fermi level (at a technical level, one should take $T\to 0$ while holding $\gamma \ll k_B T/\hbar$). The issue is that to properly resolve the depression in the transmission function at the Fermi level, one needs to properly get the Fermi level, which due to Lorentzian smearing at finite $\gamma$, does not happen. This all can be traced back to the fact that the Markovian equation does not even have a proper Fermi level. This helps shed light on the observation of Chiang and Hsu that the transmission function due to destructive interference needs to have $\gamma$ to be zero or else this peak is partially washed out. An inhomogeneous $\gamma_k$ and mode density would immediately help in this regard, as does the finite bias. The latter averages the current over some region of energy, so even if some feature is not properly obtained, one can still approximately get the current, explaining another observation of Ref.~\cite{chiang_quantum_2020} that errors get bigger as the bias gets smaller.

Chiang and Hsu also perform asymptotic analyses for their two example junctions. They first derive their Eq.~(16) for a single site junction with large, homogeneous $\gamma$ (later, they obtain the same equation, written inline, for their two site example). They then state that this ``is exactly consistent with the numerical results in the previous studies,'' citing our Refs.~\onlinecite{gruss_landauers_2016,gruss_communication_2017,elenewski_communication_2017}, as well as claim ``However, these studies do not provide any direct mathematical proof for the asymptotic behavior based on the DLvN approach.'' They further claim that, ``The previous studies have discussed the effect of electronic relaxation in leads on quantum transport for the one-level system coupled to two 1D electrodes and for a graphene nanoribbon between two gold substrates.''

However, in addition to the general solution noted above, we had already a general and direct asymptotic analysis for large $\gamma$, see Eq.~(B14), Eq.~(B17), and the large-$\gamma$ result in Eq.~(B18). {\em This result applies to arbitrary junctions} (beyond the two example junctions of Chiang and Hsu, but still requiring identical reservoirs) and also to {\em many-body systems}. Moreover, although we also took $\gamma$ to be homogeneous, this was not needed in the derivation. Allowing for inhomogeneous $\gamma$ (and complex $v_{ik}$), the result -- Eq.~(B18) of Ref.~\onlinecite{gruss_landauers_2016} -- is
\[ \label{eq:Curr_LG_PC}
I \approx 2e \sum_{k \in \ql} \sum_{i \in \qs} \frac{\left| v_{ik} \right|^2}{\gamma_k}(\tilde{f}_k^\ql-\tilde{f}_k^\qr) .
\]
This is for arbitrary noninteracting or many-body systems, but identical reservoirs, and for inhomogeneous $\gamma$. It is more general than their Eq.~(16), which applies only to two example systems where, in both cases, only one system site is connected to the reservoirs. Deriving the same result (albeit here only shown for non-interacting systems) directly from Eq.~\eqref{eq:Curr_NI_PC} yields the compact form $I \approx 2e \sum_{k \in \ql}  (\tilde{f}_k^\ql-\tilde{f}_k^\qr) \ik /\gamma_k$ with $\ik = \inner{v_k}{v_k}$.

After the large-$\gamma$ analysis, Chiang and Hsu then provide a small-$\gamma$ expression, their Eq.~(18) (and a similar expression inline for the two site model), and again state, ``This result also supports the numerical results in the previous studies," citing our work. Yet, we also derived this same equation, see Eq.~(8) in the main text of Ref.~\onlinecite{gruss_landauers_2016} (Markovian and non-Markovian relaxation have the same small-$\gamma$ behavior, as shown in the Supplemental Information there). In this case, the derivation was in linear response for identical reservoirs, and did assume that the nature of the system is irrelevant (when the small-$\gamma$ regime is reached will depend on the nature of the system, but not the form of the current in that regime). The expression is (allowing for unequal $\gamma_k$)
\[ \label{eq:Curr_WG_PC}
I \approx \frac{e}{2} \sum_{k \in \ql} \gamma_k (\tilde{f}_k^\ql-\tilde{f}_k^\qr) ,
\]
This general small-$\gamma$ result is derived from an effective approach and can benefit from analytic results and a direct derivation. This is provided by Eq.~\eqref{eq:Curr_NI_PC}. As with the other calculations above, the asymptotic results will also be generalized in the companion article~\cite{zwolak_analytic_2020}. 

Both of these regimes have the relaxation limiting transport, as discussed physically in Ref.~\onlinecite{gruss_landauers_2016} by the analogy to Kramers' problem for chemical reaction rates~\cite{kramers_brownian_1940,hanggi_reaction-rate_1990} (see also Ref.~\onlinecite{velizhanin_crossover_2015} for classical thermal transport). For small $\gamma$, the relaxation is rate-limiting and nothing about the system appears in the current (again, though, as noted above, when this regime is reached -- i.e., at what $\gamma$ -- does depend on the system). It is only the number of reservoir modes in the bias window that matters. For large-$\gamma$, the relaxation is again limiting. However, the system-reservoir couplings do appear in the large-$\gamma$ expression, Eq.~\eqref{eq:Curr_LG_PC} since in this regime the current relies on how much coherence can be generated between reservoir modes and the system before the relaxation suppresses it (i.e., it is an overdamping of the coherence needed for the current to flow -- all particle movement in quantum mechanics is associated with coherence). That being said, some alternative asymptotic regimes can appear for non-Markovian transport, see, e.g., Ref.~\onlinecite{gruss_communication_2017}. This has to do with the band structure of the reservoirs (e.g., band gaps), the level energies of the system, and symmetries. 

We conclude here by noting that we were delighted to see the interesting results obtained by Chiang and Hsu. They indeed help understand the behavior of the DLvN approach. We expect that their contribution and the more general results in Ref.~\onlinecite{gruss_landauers_2016} and here, as well future contributions, will eventually make the DLvN and its many-body analog routine in the simulation of transport. Moreover, these types of studies are necessary now more than ever due to the inroads of efficient tensor network simulations of transport~\cite{rams_breaking_2020}, specifically using many-body approaches with relaxation~\cite{wojtowicz_open-system_2020,brenes_tensor-network_2020,fugger_nonequilibrium_2020,lotem_renormalized_2020}.

\end{document}